\documentclass[12pt]{amsart}
\usepackage{amssymb, amsmath, amsthm}
\theoremstyle{plain}
\newtheorem{theorem}{Theorem}

\theoremstyle{definition}

\newcommand{\bi}{\bigtriangleup}

\newcommand{\ep}{\epsilon}
\newcommand{\Th}{\Theta}

\newcommand{\la}{\lambda}

\newcommand{\de}{\delta}

\newcommand{\pa}{\partial}

\newcommand{\no}{\nonumber}

\begin{document}
\title[Rational Symmetries of KdV]
 { \sc Rational Approximate Symmetries of KdV Equation\/}
\author[J.H. Chang]
{{ Jen-Hsu Chang \/}\\
  { Department of General Courses,\\
 Chung-Cheng Institute of Technology ,\\
   National Defense University,\/}\\
  { Dashi, Tau-Yuan County, Taiwan, 33509 }}
\date{\today \\
\indent e-mail: jhchang@ccit.edu.tw}
\maketitle
\begin{abstract}
We construct one-parameter deformation of  the Dorfman Hamiltonian operator for the Riemann hierarchy using the quasi-Miura
transformation from topological field theory. In this way, one can get the rational approximate symmetries of KdV equation and then investigate its bi-Hamiltonian structure.  
\end{abstract}

\section{Introduction}
\indent In this paper, one will investigate the one-parameter deformation  of  the Dorfman Hamiltonian 
operator ($D=\pa_x$)
\begin{equation}
J=D\frac{1}{v_x}D\frac{1}{v_x}D, \label{Do}
\end{equation}
which is third-order and compatible with the differential operator $D$, i.e., $J+\la D$ is Hamiltonian operator for any $\la$ \cite{D1}. The deformation of the bi-Hamiltonian pair $J$ and $D$ satisfies the Jacobi identity only up to a certain order of the parameter of the deformation. The problem is that how we can find the deformation such that the bi-Hamiltonian structure can be preserved. One way to construct
this deformation is borrowed from the free energy of the topological field theory(TFT) \cite{DZ}(and references therein). The free energy satisfies the universal loop equation(p.157 in \cite{DZ}). From the free energy, one can construct the so-called quasi-Miura transformation to get the deformation(see below). \\
\indent From the deformation of the bi-Hamiltonian pair, one can also get the deformation of the recursion operator $JD^{-1}$ to the genus-one correction ($\ep^2$-correction). The deformed recursion operator can be used to generate higher-order symmetries, which commute each other only up to $O(\ep^4)$. In doing so, we can deform the origional integrable system to include $\ep^2$-correction . The rational approximate  symmetries of KdV equation are established using the 
method. \\
\indent Let's start with the well-known Riemann equation
\begin{equation}
v_t=vv_x \label{Rie1}.
\end{equation}
It is also called the dispersionless KdV equation(dKdV). The integrability of \eqref{Rie1} is that it has an infinite sequence of commuting Hamiltonian flows ($t_1=t$)
\begin{equation}
v_{t_{n}}=v^n v_x, \quad n=1,\,2,\,3, \cdots \label{Rie2}.
\end{equation}
The Riemann hierarchy \eqref{Rie2} has the bi-Hamiltonian structure \cite{ON} 
\begin{equation}
v_{t_{n}}=\frac{1}{(n+1)(n+2)}D \de H_{n+2}=\frac{1}{(n+1)(n+2)(n+3)(n+4)} J \de H_{n+4}, \label{Ham}
\end{equation}
where $H_n=\int v^n dx$, $\de$ is the variational derivative and $J$ is the Dorfman 
Hamiltonian operator \eqref{Do}. From the bi-Hamiltonian structure \eqref{Ham}, the recursion operator is defined as 
\begin{equation}
L=JD^{-1}=D\frac{1}{v_x}D\frac{1}{v_x}=R^2, \label{Rec1}
\end{equation}
where 
\begin{equation}
R=D\frac{1}{v_x} \label{Rec2}
\end{equation}
is the Olver-Nutku recursion operator \cite{ON}, i.e., the square root of the recursion operator $L$.  One can easily check that $R$(or $L$) satisfies the following recursion operator equation associated with the Riemann equation \eqref{Rie1}
\begin{equation}
A_t=[v_x+vD, A], \label{Req}
\end{equation}
where $A$ is (pseudo-)differential operator.
Then from the recursion operator theory \cite{Pe}, one can establish new symmetries of 
\eqref{Rie1} by the Olver-Nutku recursion operator \eqref{Rec2} repeatedly
\begin{equation}
v_{\tau_n}=R^n 1, \quad n=1,\,2,\,3, \cdots \label{Sym}
\end{equation}
The new symmetries \eqref{Sym} of \eqref{Rie1}, i.e., 
\[(v_t)_{\tau_n}=(v_{\tau_n})_t,\]
will correspond to the "superintegrability" of the Riemann equation  \eqref{Rie1}
\cite{ST} \\
\indent Next, to deform the recursion operator \eqref{Rec2}, we use the free energy in TFT of the famous KdV equation
\begin{equation}
u_t=uu_x+ \frac{\ep^2}{12}u_{xxx} \label{KdV}
\end{equation}
to construct the quasi-Miura transformation as follows. The free energy $F$ of  KdV equation \eqref{KdV} in TFT 
has the form($F_0= \frac{1}{6}v^3$)
\[F=\frac{1}{6}v^3+\sum_{g=1}^{\infty} \ep^{2g-2} F_g(v;v_x, v_{xx}, v_{xxx}, \cdots, v^{(3g-2)}).\]
Let 
\begin{eqnarray*}
\bi F &=& \sum_{g=1}^{\infty} \epsilon^{2g-2} F_g(v;v_x, v_{xx}, v_{xxx}, \cdots, v^{(3g-2)}) \\
                 &=& F_1(v;v_x)+ \ep^2 F_2(v;v_x, v_{xx},v_{xxx},v_{xxxx}) \\
                  &+&\ep^4 F_3(v;v_x, v_{xx},v_{xxx},v_{xxxx},\cdots,v^{(7)})+\cdots .
\end{eqnarray*}
The $\bi F$ will satisfy the loop equation(p.151 in \cite{DZ})
\begin{eqnarray}
&&\sum_{r \geq 0}\frac{\pa \bi F}{\pa v^{(r)}} \pa_x^r \frac{1}{v-\lambda}+\sum_{r \geq 1} \frac{\pa \bi F}{\pa v^{(r)}} \sum_{k=1}^r \left(\begin{array}{c} r \\k \end{array} \right) \label{loop} \\
 &&\pa_x^{k-1} \frac{1}{\sqrt{v-\lambda}} \pa_x^{r-k+1} \frac{1}{\sqrt{v-\lambda}} \no \\
&=&\frac{1}{16 \lambda^2}-\frac{1}{16(v-\lambda)^2}-\frac{\kappa_0}{\lambda^2} \no \\
&+&\frac{\ep^2}{2}\sum_{k,l \geq 0}\left[\frac{\pa^2 \bi F}{\pa v^{(k)}\pa v^{(l)}}+\frac{\pa \bi F}{\pa v^{(k)}}\frac{\pa \bi F}{\pa v^{(l)}}\right] \pa_x^{k+1} \frac{1}{\sqrt{v-\lambda}}\pa_x^{l+1} \frac{1}{\sqrt{v-\lambda}} \no \\
&-&\frac{\ep^2}{16}\sum_{k \geq 0}\frac{\pa \bi F}{\pa v^{(k)}}\pa_x^{k+2} \frac{1}{(v-\lambda)^2}. \no
\end{eqnarray}  
Then we can determine $F_1, F_2, F_3, \cdots$ recursively by substituting $\bi F$ into equation \eqref{loop}. For $F_1$, one obtains 
\[\frac{1}{v-\lambda}\frac{\pa F_1}{\pa v}-\frac{3}{2}\frac{v_x}{(v-\lambda)^2}\frac{\pa F_1}{\pa v_x}=\frac{1}{16 \lambda^2}-\frac{1}{16 (v-\lambda)^2}-\frac{\kappa_0}{\lambda^2}.\]
From this, we have 
\[\kappa_0=\frac{1}{16}, \quad F_1=\frac{1}{24} \log v_x.\]
For the next terms $F_2(v;v_x, v_{xx},v_{xxx},v_{xxxx})$, it is presented in the appendix $A$. Now, one can define the quasi-Miura transformation as
\begin{eqnarray}
\label{qua}\\
 u&=&v+\ep^2 (\bi F)_{xx}=v+ \ep^2(F_1)_{xx}+ \ep^4(F_2)_{xx}+\cdots \no \\
 &=&v+\frac{\ep^2}{24}(\log v_x)_{xx}+ \ep^4(\frac{v_{xxxx}}{1152v_x^2}-\frac{7v_{xx}v_{xxx}}{1920v_x^3}+\frac{v_{xx}^3}{360v_x^4})_{xx}+ \cdots. \no
\end{eqnarray}
One remarks that Miura-type transformation means the coefficients of $\ep$ are homogeneous polynomials in the derivatives $v_x, v_{xx}, \cdots , v^{(m)}$(p.37 in \cite{DZ}, \cite{LO}) and "quasi" means the ones of $\ep$ are quasi-homogeneous rational fuctions in the derivatives, too (p.109 in \cite{DZ} and see also \cite{Sr} ). \\
\indent The truncated quasi-Miura transformation 
\begin{equation}
u=v+\sum_{n=1}^g \ep^{2n}\left[F_n(v;v_x, v_{xx}, \cdots, v^{(3g-2)}) \right]_{xx} \label{tqua}
\end{equation}
has the basic property (p.117 in \cite{DZ}) that it reduces the Magri Poisson pencil of KdV equation \eqref{KdV} \cite{MA}
\begin{equation}
\{u(x), u(y)\}_{\la}=[u(x)-\la]D\de(x-y)+\frac{1}{2}u_x(x)\de (x-y)+\frac{\ep^2}{8}D^3 \de(x-y) \label{Mag}
\end{equation}
to the Poisson pencil of  the Riemann hierarchy \eqref{Rie2}:
\begin{equation}
\{v(x), v(y)\}_{\la}=[v(x)-\la]D\de(x-y)+\frac{1}{2}v_x(x)\de (x-y)+ O(\ep^{2g+2}).\label{Mag1}
\end{equation}
One can also say that the truncated quasi-Miura transformation  \eqref{tqua} deforms the KdV equation \eqref{KdV} to the Riemann equation \eqref{Rie1} up to $O(\ep^{2g+2})$.
And conversely, we can also think that the Poisson pencil \eqref{Mag1} for the Riemann
hierarchy is deformed to get the Magri Poisson pencil \eqref{Mag} of genus-g correction after the truncated quasi-Miura transformation \eqref{tqua}. So a very natural question arises: under the truncated quasi-Miura transformation \eqref{tqua}, is the deformed 
Dorfman's Hamiltonian operator $J(\ep)$ of \eqref{Do} still Hamiltonian and compatible with $D$
up to $O(\ep^{2g+2})$ ? The answer is affirmative for $g=1$, i.e.,
\begin{equation}
u=v+ \frac{\ep^2}{24}(\log v_x)_{xx}+O(\ep^4) \label{qua1}
\end{equation}
or
\begin{equation}
v=u - \frac{\ep^2}{24}(\log u_x)_{xx}+O(\ep^4). \label{qua2}
\end{equation}
and it is the main purpose of this paper.\\
\indent   Also, from the deformed recursion operator $R(\ep)$ of the Olver-Nutku
recursion operator \eqref{Rec2}, we can generate rational approximate symmetries of KdV equation \eqref{KdV} up to $O(\ep^4)$. These symmetries are different from the ones generated 
by the Magri Poisson pencil \eqref{Mag}. Then one can call them the "superintegrability" of KdV equation .\\
\indent Finally, one remarks that in general integrable dispersive deformation for integrable dispersionless systems
is not unique \cite{CT,Ko,NU,Sr}. For deformations of bi-Hamiltonian PDEs of hydrodynamic type with one dependent variable, we refer to \cite{LO}. \\
\indent The paper is organized as follows. In the next section, we construct the genus-one deformation of Olver-Nutku recursion operator. In section 3, the bi-Hamiltonian structure 
of the rational approximate symmetries of the KdV equation \eqref{KdV} is investigated. In the final section, we discuss some problems to be investigated.

\section{Quasi-Miura Transformation of Olver-Nutku Recursion Operator} 
In this section, we will investigate the Hamiltonian operator $D$ and the Olver-Nutku
recursion operator \eqref{Rec2} under the truncated quasi-Miura transformation \eqref{tqua} for $g=1$.\\
\indent In the new "u-coordinate", $D$ and $R$ will be given by the operator 
\begin{eqnarray}
D(\ep) &=& M^* D M   \label{Td} \\
R(\ep) &=& M^* R (M^*)^{-1},  \label{Tr}
\end{eqnarray}
where
\begin{eqnarray*}
M &=& 1-\frac{\ep^2}{24}D \frac{1}{u_x}D^2 \\
M^* &=& 1+\frac{\ep^2}{24}D^2 \frac{1}{u_x}D,
\end{eqnarray*}
$M^*$ being the adjoint operator of $M$. Then using \eqref{Td}, \eqref{Tr} and \eqref{qua2},
we can yield after a simple calculation
\begin{eqnarray*}
D(\ep) &=& D+ O(\ep^4) \\
R(\ep) &=& D \frac{1}{u_x} + \frac{\ep^2}{24}D (D\frac{1}{u_x}D^2\frac{1}{u_x}-\frac{1}{u_x}D^2\frac{1}{u_x}D+ \frac{(\log u_x)_{xxx}}{u_x^2})+ O(\ep^4) \\
 &=& D \frac{1}{u_x}+ \frac{\ep^2}{12}D [\frac{1}{u_x}(\frac{1}{u_x})_x D^2+(\frac{1}{u_x}(\frac{1}{u_x})_x)_x D- 3 u_x^{-5} u_{xx}^3 \\
&+& 2 u_x^{-4} u_{xx}u_{xxx}] + O(\ep^4). 
\end{eqnarray*}
We hope that $R(\ep)$ is an recursion operator of KdV equation \eqref{KdV}. Indeed,
it's the following
\begin{theorem}
$R(\ep)$ satisfies the recursion operator equation of KdV equation
\begin{equation}
R(\ep)_t=[u_x+uD+ \frac{\ep^2}{12}D^3, R(\ep)] +O(\ep^4). \label{Req1}
\end{equation}
\end{theorem}
\begin{proof}
Direct calculations.
\end{proof}
We can think \eqref{Req1} as the genus-one deformation of \eqref{Req}. One remarks
that the recursion operator 
\[u+\frac{u_x}{2}D^{-1}+ \frac{\ep^2}{8}D^2 \] 
of Magri pencil \eqref{Mag} will also satisfy the recursion operator equation \eqref{Req1} but there is no higher-order correction. Moreover, we know that in general recursion operator is non-local \cite{Pe} and hence the local property of $R(\ep)$ is special from this point of view. \\
\indent Now from Theorem 1, one will construct infinite rational symmetries (up to $O(\ep^4)$)
 of  KdV equation \eqref{KdV} using recursion operator
$R(\ep)$ as follows:
\begin{equation}
u_{\tau_n}=R^n(\ep)1+O(\ep^4),\, n=1, \,2,\,3,\, \cdots, \label{Tau}
\end{equation}
which is the genus-one deformation of \eqref{Sym}. For example,
\begin{eqnarray}
u_{\tau_1} &=& R(\ep)  1 = [ \frac{1}{u_x} + \frac{\ep^2}{12}(- 3 u_x^{-5} u_{xx}^3+2 u_x^{-4} u_{xx}u_{xxx})]_x \label{Tau2} \\
&+& O(\ep^4)  \no \\
u_{\tau_2} &=& R^2(\ep)  1 =\{\frac{1}{u_x}(\frac{1}{u_x})_x +\frac{\ep^2}{12}[30u_x^{-7}u_{xx}^4 -  30 u_x^{-6}u_{xx}^2 u_{xxx} \no \\
&+& 3 u_x^{-5} u_{xxx}^2+3 u_x^{-5} u_{xx}u_{xxxx}] \}_x+ O(\ep^4). \no
\end{eqnarray}
\indent Also, we notice that one can also obtain \eqref{Tau} by \eqref{qua1} as follows.
Since
\[u_{\tau_{n+1}}=v_{\tau_{n+1}}+\frac{\ep^2}{24}(\frac{v_{\tau_nx}}{v_x})_{xx}+ O(\ep^4), \]
then using \eqref{qua2}, after some calculations, we can obtain
\begin{eqnarray}
u_{\tau_{n+1}}&=& \{\frac{u_{\tau_{n}}}{u_x}+ \frac{\ep^2}{24}[\frac{(\log u_x)_{xxx}}{u_x^2}u_{\tau_n}]+\frac{\ep^2}{24}[(\frac{u_{\tau_{n}}}{u_x})_{xx}/u_x]_x \no \\
&-& \frac{\ep^2}{24}[(\frac{u_{\tau_nx}}{u_x})_{xx}/u_x]\}_x +O(\ep^4) \no  \\
&=& R(\ep)u_{\tau_n}+O(\ep^4). \no
\end{eqnarray}

\section{Bi-Hamiltonian structure of rational Approximate Symmetries }
In this section, one  will prove the bi-Hamiltonian structure of \eqref{Tau} for even 
flows, i.e., $n=2k$, $k \geq 1$. \\
\indent Firstly, the deformed Dorfman Hamiltonian operator $J(\ep)$ under the quasi-Miura transformation \eqref{qua1} is
\begin{eqnarray}
J(\ep)&=&R^2(\ep)D(\ep) \label{Qon} \\
&=&D \frac{1}{u_x}D \frac{1}{u_x}D + \frac{\ep^2}{24}D[\frac{1}{u_x}D\frac{(\log u_x)_{xxx}}{u_x^2}+\frac{(\log u_x)_{xxx}}{u_x^2} \no \\
&&D\frac{1}{u_x} + D\frac{1}{u_x}D^2\frac{1}{u_x}D\frac{1}{u_x}- \frac{1}{u_x}D\frac{1}{u_x}D^2\frac{1}{u_x}D]D+ O(\ep^4). \no
\end{eqnarray}
Then we have the following 
\begin{theorem} (i) $J(\ep)$ is a Hamiltonian operator up to $O(\ep^4)$. (ii) $J(\ep)$ and $D(\ep)$ form a bi-Hamiltonian pair up to $O(\ep^4)$.
\end{theorem}
\begin{proof}(i) The skew-symmetric property of the operator \eqref{Qon} is obvious.
To prove $J(\ep)$ is Hamiltonian operator, we must verify that $J(\ep)$ satisfies
the Jacobi identities up to $O(\ep^4)$. Following \cite{Pe,ON}, we introduce the arbitrary basis of tangent vector $\Th$,  which are then conveniently manipulated 
according to the rules of exterior calculus. The Jacobi identities are given by the compact expression \begin{equation}
P(\ep) \wedge \de I= O(\ep^4) \,(mod. \, div.), \label{Van} \end{equation}
where $P(\ep)=J(\ep)\Th$, \, $I=\frac{1}{2} \Th \wedge P(\ep)$ and $\de$ denotes
the variational derivative. The vanishing of the tri-vector \eqref{Van} modulo
a divergence is equivalent to the satisfication of Jacobi identities.\\
\indent Now, a lengthy and tedious calculation can yield 
\begin{eqnarray}
P(\ep) &=&  \{\frac{1}{u_x}(\frac{\Th_x}{u_x})_x+ \frac{\ep^2}{24}[\frac{1}{u_x}(\frac{(\log u_x)_{xxx}}{u_x^2}\Th_x)_x +\frac{(\log u_x)_{xxx}}{u_x^2}(\frac{\Th_x}{u_x})_x  \no \\
&+& (\frac{1}{u_x}(\frac{1}{u_x}(\frac{\Th_x}{u_x})_x)_{xx})_x- \frac{1}{u_x}(\frac{1}{u_x}(\frac{\Th_{xx}}{u_x})_{xx})_x] \}_{x} \no  \\
&+&  O(\ep^4) \no
\end{eqnarray}
And 
\begin{eqnarray}
I&=&\frac{1}{2} \Th \wedge P(\ep)= -\frac{1}{2u_{x}^2} \Th_x \wedge \Th_{xx} \no \\
&+& \frac{\ep^2}{24} \{-5u_{x}^{-6}u_{xx}^3 \Th_x \wedge \Th_{xx}+3u_{x}^{-5}u_{xx}^2 \Th_x \wedge \Th_{xxx} \no \\
&-& 2u_{x}^{-4}u_{xx} \Th_{xx} \wedge \Th_{xxx} \}+O(\ep^4) \quad (mod. \quad  div.) \no
\end{eqnarray} 
Then 
\begin{eqnarray}
\de I&=&(3u_{x}^{-4}u_{xx} \Th_x \wedge \Th_{xx}-u_{x}^{-3} \Th_x \wedge \Th_{xxx}) \label{Del} \\
 &+& \frac{\ep^2}{24}\{ 60 u_{x}^{-7}u_{xx}^3 \Th_x \wedge \Th_{xx}-30 u_{x}^{-6}u_{xx}u_{xxx} \Th_x \wedge \Th_{xx} \no \\
&-& 30 u_{x}^{-6}u_{xx}^2 \Th_x \wedge \Th_{xxx}+ 6u_{x}^{-5}u_{xxx} \Th_x \wedge \Th_{xxx} \no \\
&+& 6u_{x}^{-5}u_{xx} \Th_{xx} \wedge \Th_{xxx}+6 u_{x}^{-5}u_{xx} \Th_x \wedge \Th_{xxxx} \no \\
&-& 2 u_{x}^{-4} \Th_{xx} \wedge \Th_{xxxx} \}_x  +O(\ep^4)\no     
\end{eqnarray}
Finally,
\begin{eqnarray}
P(\ep) \wedge \de I &=& 0-\frac{\ep^2}{24} \{-7 u_{x}^{-8}u_{xx}u_{xxx} \Th_x \wedge \Th_{xx}
\wedge \Th_{xxxx} \no \\
&+& 7 u_{x}^{-8}u_{xx}^2  \Th_x \wedge \Th_{xxx} \wedge \Th_{xxxx} \no \\
&-& 7 u_{x}^{-7}u_{xx} \Th_{xx} \wedge \Th_{xxx} \wedge \Th_{xxxx} \no \\
&+& (7 u_{x}^{-8}u_{xx}u_{xxxx}-u_{x}^{-7}u_{xxxxx}) \Th_x \wedge \Th_{xx} \wedge \Th_{xxx} \no \\
&-& u_{x}^{-7}u_{xx} \Th_x \wedge \Th_{xxx} \wedge \Th_{xxxxx} \no \\
&+& u_{x}^{-7}u_{xxx} \Th_x \wedge \Th_{xx} \wedge \Th_{xxxxx} \no \\
&+&  u_{x}^{-6} \Th_{xx} \wedge \Th_{xxx} \wedge \Th_{xxxxx} \} \no \\
&=& (u_{x}^{-6} \Th_{xx} \wedge \Th_{xxx}\wedge \Th_{xxxx})_x \no \\
&-& (u_{x}^{-7} u_{xx} \Th_{x} \wedge \Th_{xxx}\wedge \Th_{xxxx})_x \no \\
&+& (u_{x}^{-7} u_{xxx} \Th_{x} \wedge \Th_{xx}\wedge \Th_{xxxx})_x \no \\
&-& (u_{x}^{-7} u_{xxxx} \Th_{x} \wedge \Th_{xx}\wedge \Th_{xxx})_x+O(\ep^4) \no
\end{eqnarray} 
which is a total derivative so that the Jacobi identities are satisfied and this complete
the proof of $(i)$.\\
(ii) Since $J(\ep)$ and $D(\ep)$ are Hamiltonian operators,  we need only verify the additional 
condition 
\[P(\ep) \wedge \de I_D + D(\Th) \wedge \de I=O(\ep^4),\]
where 
\[I_D=\frac{1}{2} \Th \wedge D (\Th) =\frac{1}{2} \Th \wedge \Th_x,\]
$\de I$ and  $P(\ep)$ are defined in \eqref{Van}, modulo a divergence.\\
\indent Obviously, $\de I_D= O(\ep^4)$. So we will check $D(\Th) \wedge \de I=\Th_x \wedge
 \de I =O(\ep^4)$. From \eqref{Del}, we have 
\begin{eqnarray}
\Th_x \wedge \de I &=& 0+ \frac{\ep^2}{24} \{6 u_{x}^{-5} u_{xxx} \Th_{x} \wedge \Th_{xx}\wedge \Th_{xxx} \no \\
&-& 30 u_{x}^{-6} u_{xx}^2 \Th_{x} \wedge \Th_{xx}\wedge \Th_{xxx} \no \\
&+& 6 u_{x}^{-5} u_{xx} \Th_{x} \wedge \Th_{xx}\wedge \Th_{xxxx} \}  \no \\
&=& \frac{\ep^2}{24} \{6 u_{x}^{-5} u_{xx} \Th_{x} \wedge \Th_{xx}\wedge \Th_{xxx} \}_{x}
+ O(\ep^4) \no 
\end{eqnarray}
This completes the proof of $(ii)$.
\end{proof}
{\bf Remark:} Although the quasi-Miura transformation \eqref{qua1} is  of change of coordinates (including derivatives), it is non-trivial to see that $J(\ep)$ is a Hamiltonian operator (up to $O(\ep^4)$). It's because that change of coordinates, in general, won't preserve the Jacobi identities.\\
\indent Since $J(\ep)$ and $D(\ep)$ form a Hamiltonian pair, we will find the Hamiltonian densities of the even flows of the rational approximate symmetries of KdV equation \eqref{KdV} up to $O(\ep^4)$
\begin{equation} 
u_{\tau_{2n}}=R^{2n}(\ep)1= D(\ep) \frac{\de \tilde H_n(\ep)}{\de u}=J(\ep)\frac{\de \tilde H_{n-1}(\ep)}{\de u},\quad  n=1, \quad 2, \quad 3 \cdots,  \label{Even}
\end{equation}
in the following way. Firstly, we notice that by \eqref{Sym} we have 
\begin{equation}
v_{\tau_{2n}}=R^{2n}1=D(K_{2n+1})=J(K_{2n-1}), \label{Even2} 
\end{equation}
where
\begin{eqnarray*}
K_1 &=& x  \\
K_3 &=& \frac{1}{v_x}( \frac{1}{v_x})_x \\
K_5 &=& \frac{1}{v_x}( \frac{1}{v_x}(K_3)_x)_x \\
    &\vdots& \\
K_{2n+1} &=& \frac{1}{v_x}( \frac{1}{v_x}(K_{2n-1})_x)_x. \\
\end{eqnarray*}
From the bi-Hamiltonian structure of $J$ and $D$, one can construct the Hamiltonian densities
of \eqref{Even2} using the method described in \cite{ID}. Secondly, by the Hamiltonian 
structure of $J(\ep)$ and $D(\ep)$, one can also construct the Hamiltonian densities of \eqref{Even}
using the quasi-Miura transformation \eqref{qua1}. For example,
\begin{eqnarray*}
v_{\tau_{2}} &=& R^{2}1=D(\frac{1}{v_x}( \frac{1}{v_x})_x)=D\frac{\de \hat H_1}{\de v}=J\frac{\de \hat H_0}{\de v}  \\
v_{\tau_{4}} &=& R^{4}1=D(\frac{1}{v_x}( \frac{1}{v_x}(K_3)_x)_x)=D\frac{\de \hat H_2}{\de v}=J\frac{\de \hat H_1}{\de v},  \\
\end{eqnarray*}
where
\begin{eqnarray*}
\hat H_0 &=& \int xv dx \\
\hat H_1 &=& \frac{1}{2} \int \frac{1}{v_x} dx \\
\hat H_2 &=& -\frac{1}{2} \int v_{xx}^2 v_x^{-5}  dx. \\
\end{eqnarray*}
Then after the quasi-Miura transformation \eqref{qua2}, one can obtain 
\begin{eqnarray*}
\hat H_0(\ep) &=& \int x(u-\frac{\ep^2}{24}(\log u_x)_{xx}) dx + O(\ep^4) \\
\hat H_1(\ep) &=& \frac{1}{2} \int [\frac{1}{u_x} +\frac{\ep^2}{24}(2u_x^{-5}u_{xx}^3-3u_x^
{-4}u_{xx}u_{xxx}+u_x^{-3}u_{xxxx})] dx \\
&+&  O(\ep^4) \\
\hat H_2(\ep) &=& -\frac{1}{2} \int [u_{xx}^2 u_x^{-5}+ \frac{\ep^2}{24}( 22u_x^{-9}u_{xx}^5-39u_x^{-8}u_{xx}^3u_{xxx} \\
&+&13u_x^{-7}u_{xx}^2 u_{xxxx}- 2u_x^{-6}u_{xx} u_{xxxxx}+ 6u_x^{-7}u_{xx} u_{xxx}^2)] dx \\
&+&  O(\ep^4). \\                      
\end{eqnarray*}
On the other hand, we can also verify using MAPLE that, noting \eqref{Tau2},
\begin{eqnarray*}
u_{\tau_2} &=& R^2(\ep) 1=D(\ep)\frac{\de \hat H_1(\ep)}{\de u}=J(\ep)\frac{\de \hat H_0(\ep)}{\de u}  + O(\ep^4) \\
u_{\tau_4} &=& R^4(\ep) 1=\{\frac{1}{u_x}( \frac{1}{u_x}(\frac{1}{u_x})_x)_x+\frac{\ep^2}{12}
[ 1050 u_x^{-9}u_{xx}^3 u_{xxxx}-105 u_x^{-8}u_{xxx}^3 \\
&+& 3780 u_x^{-11}u_{xx}^6 - 6300 u_x^{-10}u_{xx}^4 u_{xxx}+ 2310 u_x^{-9}u_{xx}^2 u_{xxx}^2\\
&-&420u_x^{-8}u_{xx}u_{xxx}u_{xxxx}+ 5u_x^{-7}u_{xx} u_{xxxxxx} + 15 u_x^{-7}u_{xxx} u_{xxxxx}\\
&+& 10u_x^{-7} u_{xxxx}^2-105u_x^{-8}u_{xx}^2 u_{xxxxx}]\}_x +  O(\ep^4) \\
&=& D(\ep)\frac{\de \hat H_2(\ep)}{\de u}=J(\ep)\frac{\de \hat H_1(\ep)}{\de u},
\end{eqnarray*}
which comes from the fact that the quasi-Miura transformation for $g=1$ is canonical by theorem 2. \\
\indent One remarks that the truncated  ${\tau_{2n}}$-flows are approximately integrable systems. We expect that solutions to such approximately integrable equations exhibit an integrable behavior at least for small physical parameters, for example, solitons solutions, as in \cite{LO}. But the truncated  $\tau_{2n}$-flows  are very complicated and need 
further investigations.

\section{Concluding Remarks}
We have studied the genus-one deformation of Dorfman Hamiltonian operator using quasi-Miura
transformation borrowed from the free energy of the topological field theory. Then one can prove that the deformed Hamiltonian operators $J(\ep)$ and $D(\ep)$ are still compatible and thus it provides 
the rational approximate symmetries of the KdV equation up to $O(\ep^4)$. \\
\indent In spite of the results obtained, there are some interesting issues deserving 
more investigations:
\begin{itemize}
\item We believe that Theorem 1 and Theorem 2 can be generalized to higher genus, i.e., $g \geq 2$. However, the computations will become quite unmanageable. \\
\item The Schwarzian KdV equation ( degenerate Krichever-Novikov(KN) equation \cite{KN} or Ur-KdV equation \cite{GW}) is
\[ v_t=v_{xxx}-\frac{3}{2} v_x^{-1}v_{xx}^2=v_x \{v, x\},\]
where $\{v,x\}$ is the Schwarzian derivative. It is known that 
\[\frac{1}{v_x}D \frac{1}{v_x}\] 
and the Dorfman Hamiltonian operator $J$  constitute a symplectic pair of the Schwarzian KdV equation \cite{ID}.
Thus under the quasi-Miura transformation we can also investigate the genus-one deformation
of the Schwarzian KdV equation \cite{CH}. \\
\item One can generalize $J$ to the polytropic gas system \cite{ON}. Using the universal loop equation of free energy (p.157 in \cite{DZ}), we can also find the corresponding quasi-Miura transformations of two variables and study their deformations. Thus the rational approximate symmetries of polytropic gas systems will also be obtained. But the computations are more involved and need further investigations.\\

\end{itemize}
{\bf Acknowledgments\/} \\
The author is grateful for Prof. Nutku's stimulating conversations  on the Hamiltonian theory
of polytropic gas system. He would like to thank the unknown referees for their valuable suggestions. He also thanks for National Science Council under grant no. NSC 91-2115-M-014-001 for support.  
\newpage
\appendix
\section{}
The equation for $F_2$ is
\begin{eqnarray*}
&&\frac{1}{(v-\lambda)^5}(\frac{105}{2048}v_x^2-\frac{945}{16}v_x^4 \frac{\pa F_2}{\pa v_{xxxx}}) \\
&&+\frac{1}{(v-\lambda)^4}(\frac{-49}{1536}v_{xx}+\frac{735}{8}v_x^2 v_{xx} \frac{\pa F_2}{\pa v_{xxxx}}+\frac{105}{8}v_x^3  \frac{\pa F_2}{\pa v_{xxx}} ) \\
&&+\frac{1}{(v-\lambda)^3}[\frac{v_{xxx}}{192v_x}-\frac{23 v_{xx}^2}{4608v_x^2}-(16v_{xx}^2 +\frac{87}{4}v_xv_{xxx}) \frac{\pa F_2}{\pa v_{xxxx}} \\
&&-\frac{55}{4}v_xv_{xx}\frac{\pa F_2}{\pa v_{xxx}} -\frac{15}{4}v_x^2\frac{\pa F_2}{\pa v_{xx}}] \\
&&+\frac{1}{(v-\lambda)^2}(3v_{xxxx}\frac{\pa F_2}{\pa v_{xxxx}}+\frac{5}{2}v_{xxx}\frac{\pa F_2}{\pa v_{xxx}}+2v_{xx}\frac{\pa F_2}{\pa v_{xx}}+\frac{3}{2}v_x \frac{\pa F_2}{\pa v_x}) \\
&&-\frac{1}{(v-\lambda)}\frac{\pa F_2}{\pa v}=0.
\end{eqnarray*}
Let the coefficients of $\frac{1}{(v-\lambda)^i}$, $i=1,2,3,4,5$, be equal to zero. Then one can obtain 
\[F_2=\frac{v_{xxxx}}{1152v_x^2}-\frac{7v_{xx}v_{xxx}}{1920v_x^3}+\frac{v_{xx}^3}{360v_x^4}.\]

\end{document}